\documentclass[aps,prb,twocolumn,superscriptaddress,floatfix,letter]{revtex4-1}
\usepackage{epsfig,amsmath,amssymb,color,dsfont}
\usepackage[bookmarks=true,colorlinks,linkcolor=blue,urlcolor=blue,citecolor=blue]{hyperref}
\usepackage{bm}
\usepackage{booktabs}
\usepackage{array}

\makeatletter
\newcommand{\thickhline}{%
    \noalign {\ifnum 0=`}\fi \hrule height 1pt
    \futurelet \reserved@a \@xhline
}
\newcolumntype{"}{@{\hskip\tabcolsep\vrule width 1pt\hskip\tabcolsep}}
\makeatother
\begin{document}

\title{Spin structure factors of chiral quantum spin liquids on the kagome lattice}
\author{Jad C. Halimeh}
\affiliation{Physics Department and Arnold Sommerfeld Center for Theoretical Physics, Ludwig-Maximilians-Universit\"at M\"unchen, D-80333 M\"unchen, Germany}

\author{Matthias Punk}
\affiliation{Physics Department and Arnold Sommerfeld Center for Theoretical Physics, Ludwig-Maximilians-Universit\"at M\"unchen, D-80333 M\"unchen, Germany}
\affiliation{Center for NanoScience, Ludwig-Maximilians-Universit\"at M\"unchen, D-80333 M\"unchen, Germany}

\date{\today}

\begin{abstract}
We calculate dynamical spin structure factors for gapped chiral spin liquid states in the spin-1/2 Heisenberg antiferromagnet on the kagome lattice using Schwinger-boson mean-field theory. In contrast to static (equal-time) structure factors, the dynamical structure factor shows clear signatures of time-reversal symmetry breaking for chiral spin liquid states. In particular, momentum inversion $\bm{k} \to -\bm{k}$ symmetry as well as the six-fold rotation symmetry around the $\Gamma$-point are lost. We highlight other interesting features, such as a relatively flat onset of the two-spinon continuum for the \emph{cuboc1} state. Our work is based on the projective symmetry group classification of time-reversal symmetry breaking Schwinger-boson mean-field states by Messio, Lhuillier, and Misguich.
\end{abstract}

\maketitle

\section{Introduction}

The potential to realize interesting quantum spin liquid states with fractionalized excitations and topological order has driven research on frustrated magnets in the last decades.\cite{Balents2010, Sachdev2008, Savary, Norman} One of the most promising candidate models is the spin-1/2 Heisenberg antiferromagnet on the two-dimensional kagome lattice. Many theoretical attempts have been made to unravel its groundstate properties, which are still not fully understood. While early approaches supported a symmetry broken valence bond solid state,\cite{Marston,Elser1993} various different groundstates have been proposed since. Recent numerical works based on the density matrix renormailzation group (DMRG) method provide strong evidence for a gapped $\mathbb{Z}_2$ spin liquid state,\cite{White2011, HongChen, Schollwoeck2012,Alba2015} whereas projected wavefunction studies favor a gapless $U(1)$-Dirac spin liquid groundstate,\cite{Ran, Iqbal1, Iqbal2} but this issue is not settled yet.\cite{Li2016} Both of these states do not break lattice symmetries and lack conventional long-range magnetic order due to strong quantum fluctuations associated with the frustrated spin-exchange interactions.

The interest in chiral spin liquids, which break time-reversal and parity symmetries, was triggered by Kalmeyer and Laughlin, who proposed that bosonic analogues of fractional quantum Hall states could be realized in frustrated magnets.\cite{Kalmeyer} Within a slave-fermion approach these chiral states are stable phases of matter, because gauge fluctuations are gapped by a Chern-Simons term.\cite{WenWilczekZee} More recently, various theoretical works showed that such chiral spin liquids can be stabilized on the kagome lattice either by including further-neighbor interactions or additional terms that explicitly break time-reversal symmetry.\cite{Thomale, Sheng0, Sheng1, Sheng2, Bauer, Chen, Hu, Bieri, Kumar, Wietek}

As far as experiments are concerned, the mineral Herbertsmithite as well as organic charge transfer salts are the most promising candidate materials to host a spin liquid groundstate.\cite{Nocera2005,  Helton2007, Harrison2009, McKenzie} While measurements on the triangular lattice organic salts are consistent with a gapless spin liquid, the kagome lattice compound Herbertsmithite likely has a gapped spin liquid groundstate. Inelastic neutron scattering experiments are compatible with a continuum of fractionalized spinon excitations,\cite{Lee2012} and recent NMR measurements indicate that the groundstate is gapped.\cite{Fu} The fact that no sharp onset of the two-spinon continuum was observed in neutron scattering has been attributed to the presence of a flat band of topological vison excitations in gapped $\mathbb{Z}_2$ spin liquids,\cite{Punk2014} as well as to the contribution from impurities at low energies.\cite{Han2}  

Various different spin liquid states have been proposed as potential groundstates of kagome Heisenberg antiferromagnets. In order to relate theoretical results to inelastic neutron scattering experiments, a better characterization of dynamical structure factors in kagome systems is clearly beneficial. In this work we take a step in this direction by computing dynamical spin structure factors of simple chiral spin liquids using Schwinger-boson mean-field theory.\cite{Sachdev1992, Wang} Our approach is based on an earlier projective symmetry group classification of time-reversal symmetry breaking mean-field ans\"atze by Messio, Lhuillier, and Misguich.\cite{Misguich2013}
We show that the dynamical spin structure factor $S(\bm{k},\omega)$ shows clear signatures of time-reversal symmetry breaking, in contrast to static (equal-time) structure factors. In particular, momentum inversion symmetry $\bm{k} \rightarrow -\bm{k}$ is lost and consequently the six-fold rotation symmetry of $S(\bm{k},\omega)$ around the $\Gamma$-point is reduced to three-fold rotations. Moreover, we show that the onset of the two-spinon continuum is rather flat for the \emph{cuboc1} state, which has been argued to minimize the groundstate energy of the kagome Heisenberg antiferromagnet within the Schwinger-boson approach.\cite{Lhuillier2012} This particular chiral spin liquid state is a quantum disordered version of the magnetically ordered \emph{cuboc1} state, which is a possible non-coplanar state of the classical AFKM model. \cite{Janson, Misguich2011} 

It is important to note that this Schwinger-boson construction does not lead to chiral spin liquids of the Kalmeyer-Laughlin type. This is due to the fact that the condensation of boson bilinears reduces the gauge symmetry from $U(1)$ to $\mathbb{Z}_2$. Consequently, the effective low energy theory is a Chern-Simons-Higgs theory with a condensed charge-2 Higgs field, the topological properties of which are typically equivalent to $\mathbb{Z}_2$ gauge theory.\cite{Witten, Barkeshli}

The remainder of our paper is structured as follows: In section~\ref{sec:diag} we review the Schwinger-boson mean-field theory (SBMFT) for time-reversal symmetry breaking ans\"atze and calculate the spinon dispersions. In section~\ref{sec:self} we determine the mean-field parameters self-consistently for all SBMFT ans\"atze considered in this work.  In section~\ref{sec:ssf} the spin structure factor for a general chiral SBMFT ansatz is derived. Lastly, in section~\ref{sec:results}  we present and discuss the numerically computed structure factors. We conclude with section~\ref{sec:conclusion}.

\section{Model and Methods}\label{sec:diag}

The Hamiltonian of the antiferromagnetic Heisenberg model is given by
\begin{equation}
\hat{H} = J\sum_{\langle lj\rangle}\mathbf{\hat{S}}_l\cdot\mathbf{\hat{S}}_j,
\end{equation}
where $J>0$, $\mathbf{\hat{S}}_l$ is the spin operator on lattice site $l$ and the sum runs over nearest neighbor sites. In the following we use the Schwinger-boson representation of spin operators
\begin{equation}\label{}
\mathbf{\hat{S}}_l=\frac{1}{2}\hat{b}_{l\alpha}^{\dagger}\bm{\hat{\sigma}}_{\alpha\beta}\hat{b}_{l\beta},
\end{equation}
where here, and throughout this paper, we employ a summation convention over repeated Greek indices, and $\hat{b}_{l\alpha}$, $\hat{b}_{l\alpha}^{\dagger}$ are bosonic annihilation and creation operators, respectively, of spin $\alpha$ on site $l$. 
Consequently, the Hamiltonian can be written as
\begin{align}
\nonumber \hat{H}&=\frac{J}{4}\sum_{\langle lj\rangle}\left(2\delta_{\alpha\mu}\delta_{\beta\gamma}-\delta_{\alpha\beta}\delta_{\gamma\mu}\right)\hat{b}_{l\alpha}^{\dagger}\hat{b}_{l\beta}\hat{b}_{j\gamma}^{\dagger}\hat{b}_{j\mu} \\
&+\lambda\sum_j\left(\hat{n}_j-2\mathcal{S}\right),
\end{align}
where the Langrange multiplier $\lambda$ constrains the number of bosons per site to $2\mathcal{S}$, with $\mathcal{S}$ the length of the spin. Note that this length constraint is only imposed on average here and in the following mean-field approximation. It can be enforced exactly by allowing for a space- and imaginary time dependent Lagrange multiplier, which leads to a theory of bosonic spinons coupled to an emergent U(1) gauge field.\cite{Sachdev1992} However, for the spin liquid states considered in this work, the condensation of bosonic bilinears gaps out gauge fluctuations and the mean-field approximation is justified.

\subsection{Schwinger-Boson Mean-Field Theory}

\begin{figure}
\includegraphics[width=0.95 \columnwidth]{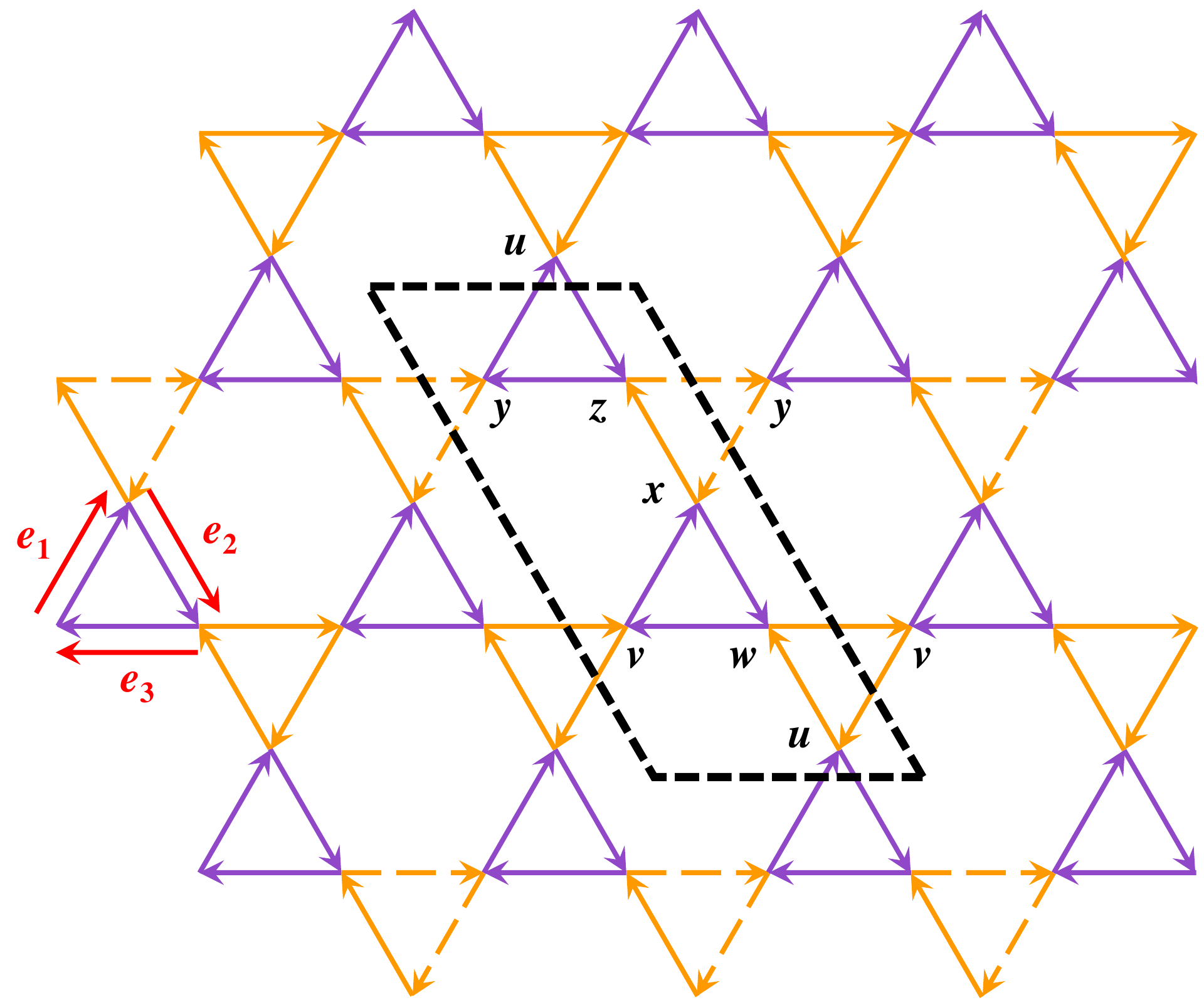}
\caption{(Color online) The six-site unit cell of the general chiral SBMFT ansatz as discussed in Refs.~[\onlinecite{Lhuillier2012}, \onlinecite{Misguich2013}]. The bond operators $\langle \hat{\mathcal{A}}_{lj}\rangle = |\hat{\mathcal{A}}_{lj}|e^{\text{i}\theta_{\mathcal{A}}}$ and $\langle \hat{\mathcal{B}}_{lj}\rangle = |\hat{\mathcal{B}}_{lj}|e^{\text{i}\theta_{\mathcal{B}}}$ between two neighboring sites $l$ and $j$ are such that at every bond one has $|\hat{\mathcal{A}}_{lj}|=\mathcal{A}$ and $|\hat{\mathcal{B}}_{lj}|=\mathcal{B}$. On purple (dark) bonds $\theta_{\mathcal{A}}=0$ and $\theta_{\mathcal{B}}=\phi_{\mathcal{B}}$, while on orange (bright) bonds the phases are $\theta_{\mathcal{A}}=\phi_{\mathcal{A}'}+\varphi$ and $\theta_{\mathcal{B}}=\phi_{\mathcal{B}'}+\varphi$ with $\varphi=0$ on undashed bonds and $\varphi=p_1\pi$ on dashed bonds, where $p_1 \in \{ 0, 1 \}$ depending on the ansatz. Finally, the red arrows indicate the real-space vectors $\bm{e}_1=a(1/2,\sqrt{3}/2)$, $\bm{e}_2=a(1/2,-\sqrt{3}/2)$, and $\bm{e}_3=a(-1,0)$, with $a$ the spacing between two neighboring sites, and $k_j=\bm{k}\cdot\bm{e}_j$.}
\label{Fig:unitCell}
\end{figure}

We now introduce the $SU(2)$-invariant bond operators
\begin{align}
\hat{\mathcal{A}}_{lj}&=\frac{1}{2}\varepsilon^{\alpha\beta}\hat{b}_{l\alpha}\hat{b}_{j\beta},\\
\hat{\mathcal{B}}_{lj}&=\frac{1}{2}\hat{b}_{l\alpha}^{\dagger}\hat{b}_{j\alpha},
\end{align}
where $\varepsilon^{\alpha\beta}$ is the fully antisymmetric tensor of $SU(2)$. One can show that $\mathbf{\hat{S}}_l\cdot\mathbf{\hat{S}}_j=\left(\hat{\mathcal{B}}^{\dagger}_{lj}\hat{\mathcal{B}}_{lj}-\hat{\mathcal{A}}^{\dagger}_{lj}\hat{\mathcal{A}}_{lj}\right)$ for $l \neq j$ and the Hamiltonian can be rewritten as
\begin{equation}
\hat{H}=J\sum_{\langle lj\rangle}\left(\hat{\mathcal{B}}^{\dagger}_{lj}\hat{\mathcal{B}}_{lj}-\hat{\mathcal{A}}^{\dagger}_{lj}\hat{\mathcal{A}}_{lj}\right)+\lambda\sum_j\left(\hat{n}_j-2\mathcal{S}\right).
\end{equation}
Next, we apply a mean-field decoupling of the bond operators 
resulting in the mean-field Hamiltonian
\begin{align}
\nonumber \hat{H}_{\text{MF}}&=J\sum_{\langle lj\rangle}\left(\langle \hat{\mathcal{B}}_{lj}\rangle\hat{\mathcal{B}}_{lj}^{\dagger}-\langle \hat{\mathcal{A}}_{lj}\rangle\hat{\mathcal{A}}_{lj}^{\dagger}+\text{H.c.}\right)\\ \nonumber
&+J\sum_{\langle lj \rangle}\left(\langle \hat{\mathcal{A}}_{lj}^{\dagger}\rangle\langle\hat{\mathcal{A}}_{lj}\rangle-\langle \hat{\mathcal{B}}_{lj}^{\dagger}\rangle\langle\hat{\mathcal{B}}_{lj}\rangle\right)\\
&+\lambda\sum_j\left(\hat{n}_j-2\mathcal{S}\right).
\end{align}
$\langle \mathcal{A}_{lj}\rangle$ and $\langle \mathcal{B}_{lj}\rangle$ are free complex mean-field parameters that will be computed self-consistently by extremizing the free energy. Even though most SBMFT studies use one or the other, including both $\langle \mathcal{A}_{lj}\rangle$ and $\langle \mathcal{B}_{lj}\rangle$ has been proven to lead to a better description of the spectrum of excitations in frustrated magnets,\cite{Trumper2011,Coleman2009} where $\langle \mathcal{A}_{lj}\rangle$ describes singlet amplitudes and $\langle \mathcal{B}_{lj}\rangle$ describes boson hopping amplitudes. A set $\{\langle \mathcal{A}_{lj}\rangle$,$\langle \mathcal{B}_{lj}\rangle\}$ specifies a mean-field ansatz. For the symmetric, time-reversal breaking spin-liquids considered in Refs.~[\onlinecite{Lhuillier2012}, \onlinecite{Misguich2013}] the mean-field parameters take the form
\begin{align}
\langle \hat{\mathcal{A}}_{lj}\rangle &=\langle \hat{\mathcal{A}}^{\dagger}_{lj}\rangle^*= |\hat{\mathcal{A}}_{lj}|e^{\text{i}\theta_{\mathcal{A}}}, \\
\langle \hat{\mathcal{B}}_{lj}\rangle &=\langle \hat{\mathcal{B}}^{\dagger}_{lj}\rangle^*= |\hat{\mathcal{B}}_{lj}|e^{\text{i}\theta_{\mathcal{B}}}, 
\end{align} 
where the moduli $|\hat{\mathcal{A}}_{lj}|=\mathcal{A}$ and $|\hat{\mathcal{B}}_{lj}|=\mathcal{B}$ are the same on each bond, but the phases $\theta_{\mathcal{B}}$ and $\theta_{\mathcal{A}}$ are bond-dependent. The detailed form of these ans\"atze is shown in Fig.~\ref{Fig:unitCell}.  Taking the Fourier transform of the Schwinger-boson operator as
\begin{equation}
\hat{b}^s_{l\alpha}=\frac{1}{\sqrt{N_{\bm{q}}}}\sum_{\bm{q}}\hat{b}^s_{\bm{q}\alpha}e^{\text{i}\bm{q}\cdot\bm{r}_l},
\end{equation}
where $s$ is a band index, $\bm{r}_l$ is the position of site $l$, and $N_{\bm{q}}$ is the number of $\bm{q}$-points summed over in the Fourier transform, and adopting a general chiral ansatz following the notation of Ref.~[\onlinecite{Misguich2013}], the Hamiltonian in reciprocal space reads
\begin{align}\label{eq:MFH}
\nonumber \hat{H}_{\text{MF}}&=\sum_{\bm{k}}\hat{\Psi}^{\dagger}_{\bm{k}}D_{\bm{k}}\hat{\Psi}_{\bm{k}}\\
&+2JN_s\left(|\mathcal{A}|^2-|\mathcal{B}|^2\right)-\lambda N_s(1+2\mathcal{S}),
\end{align}
where we have introduced the spinor 

%\begin{equation}
%\hat{\Psi}_{\bm{k}} = \left(\hat{b}^{u}_{\bm{k}\uparrow}\dots\hat{b}^{z}_{\bm{k}\uparrow},\hat{b}^{u%\dagger}_{-\bm{k}\downarrow}\dots\hat{b}^{z\dagger}_{-\bm{k}\downarrow}\right)^{\intercal}, 
%\end{equation}

\begingroup
\renewcommand*{\arraystretch}{1.5}
\begin{equation}
\hat{\Psi}_{\bm{k}} = \begin{pmatrix}
		\hat{b}^{u}_{\bm{k}\uparrow} \\
		\vdots \\
		\hat{b}^{z}_{\bm{k}\uparrow} \\
		\hat{b}^{u\dagger}_{-\bm{k}\downarrow} \\
		\vdots \\
		\hat{b}^{z\dagger}_{-\bm{k}\downarrow}
		\end{pmatrix},
\end{equation}
\endgroup

\noindent with the superscript letters denoting one of the six bands $\{u,v,w,x,y,z\}$ of the six-site unit-cell shown in Fig.~\ref{Fig:unitCell}, and $D_{\bm{k}}$ is the Hermitian block matrix
\begin{equation}
\label{eq:D}
D_{\bm{k}} = 
	\lambda\mathds{1}_{12}+C_{\bm{k}},
\end{equation}
with $\mathds{1}_d$ the $d\times d$ identity matrix, and

\sbox1{$\begin{matrix}\end{matrix}$}

\sbox2{$\begin{matrix}P_{\bm{k}}(\phi_{\mathcal{A}'})&Q_{\bm{k}}\\Q_{\bm{k}}&P_{\bm{k}}(\phi_{\mathcal{A}'}+p_1\pi)\end{matrix}$}

\sbox3{$\begin{matrix}\\\end{matrix}$}

\begin{widetext}
\begin{equation}
\label{eq:C}
C_{\bm{k}}=\frac{J}{2}\begin{pmatrix}
	R_{\bm{k}}(\phi_{\mathcal{B}'})&T_{\bm{k}} &P_{\bm{k}}(\phi_{\mathcal{A}'})&Q_{\bm{k}} \\
	T^{\dagger}_{\bm{k}}&R_{\bm{k}}(\phi_{\mathcal{B}'}+p_1\pi) & Q_{\bm{k}}&P_{\bm{k}}(\phi_{\mathcal{A}'}+p_1\pi) \\
	P^{\dagger}_{\bm{k}}(\phi_{\mathcal{A}'})&Q^{\dagger}_{\bm{k}} &R^*_{-\bm{k}}(\phi_{\mathcal{B}'})&T^*_{-\bm{k}}\\
	Q^{\dagger}_{\bm{k}}&P^{\dagger}_{\bm{k}}(\phi_{\mathcal{A}'}+p_1\pi)&T^{\intercal}_{-\bm{k}}&R^*_{-\bm{k}}(\phi_{\mathcal{B}'}+p_1\pi)
	\end{pmatrix},
\end{equation}
\begin{align}
\label{eq:R}
R_{\bm{k}}(\nu)=\mathcal{B}&\begin{pmatrix}
0 & e^{-\text{i}(\nu-k_1)} & e^{\text{i}(\phi_{\mathcal{B}'}-k_2)}\\
e^{\text{i}(\nu-k_1)} & 0 & e^{-\text{i}(\nu-k_3)}+e^{-\text{i}(\phi_{\mathcal{B}}+k_3)}  \\
e^{-\text{i}(\phi_{\mathcal{B}'}-k_2)} & e^{\text{i}(\nu-k_3)}+e^{\text{i}(\phi_{\mathcal{B}}+k_3)} & 0 
\end{pmatrix}, \\
\label{eq:T}
T_{\bm{k}}=
\mathcal{B}&\begin{pmatrix}
0 & e^{-\text{i}(\phi_{\mathcal{B}}+k_1)} & e^{\text{i}(\phi_{\mathcal{B}}+k_2)} \\
e^{\text{i}(\phi_{\mathcal{B}}+k_1)} & 0 & 0 \\
e^{-\text{i}(\phi_{\mathcal{B}}+k_2)} & 0 & 0
\end{pmatrix},\\
\label{eq:P}
P_{\bm{k}}(\nu)=\mathcal{A}&\begin{pmatrix}
0 & -e^{\text{i}(\nu+k_1)} & e^{\text{i}(\phi_{\mathcal{A}'}-k_2)} \\
e^{\text{i}(\nu-k_1)} & 0 & -e^{\text{i}(\nu+k_3)}-e^{-\text{i}k_3} \\
-e^{\text{i}(\phi_{\mathcal{A}'}+k_2)} & e^{\text{i}(\nu-k_3)}+e^{\text{i}k_3} & 0 
\end{pmatrix},\\
\label{eq:Q}
Q_{\bm{k}}=\mathcal{A}&\begin{pmatrix}
0 & -e^{-\text{i}k_1} & e^{\text{i}k_2} \\
e^{\text{i}k_1} & 0 & 0 \\
-e^{-\text{i}k_2} & 0 & 0
\end{pmatrix}.
\end{align}
\end{widetext}

\noindent Here we denote the real-space vectors $\bm{e}_1=a(1/2,\sqrt{3}/2)$, $\bm{e}_2=a(1/2,-\sqrt{3}/2)$, and $\bm{e}_3=a(-1,0)$, and $k_j=\bm{k}\cdot\bm{e}_j$, as shown in Fig.~\ref{Fig:unitCell}, with $j\in\{1,2,3\}$ and $a$ the inter-site spacing, which we set to unity.

\subsection{Bogoliubov transformation}
Finally we perform a Bogoliubov transformation by defining the bosonic operators 

%\begin{equation}
%\hat{\Gamma}_{\bm{k}}=\left(\hat{\gamma}^{u}_{\bm{k}\uparrow}\dots\hat{\gamma}^{z}_{\bm{k}%%\uparrow},\hat{\gamma}^{u\dagger}_{-\bm{k}\downarrow}\dots\hat{\gamma}^{z\dagger}_{-\bm{k}%\downarrow}\right)^{\intercal},
%\end{equation}

\begingroup
\renewcommand*{\arraystretch}{1.5}
\begin{equation}
\hat{\Gamma}_{\bm{k}}=\begin{pmatrix}
		\hat{\gamma}^{u}_{\bm{k}\uparrow} \\
		\vdots \\
		\hat{\gamma}^{z}_{\bm{k}\uparrow} \\
		\hat{\gamma}^{u\dagger}_{-\bm{k}\downarrow} \\
		\vdots \\
		\hat{\gamma}^{z\dagger}_{-\bm{k}\downarrow}
		\end{pmatrix}
\end{equation}
\endgroup

\noindent which are related to the Schwinger-boson ladder operators by the linear transformation 
\begin{equation}
\hat{\Psi}_{\bm{k}}=M_{\bm{k}}\hat{\Gamma}_{\bm{k}},
\end{equation}
whereby $\hat{\Gamma}$ will satisfy the canonical commutation relations and diagonallze $\hat{H}_{\text{MF}}$ if
\begin{align}
M^{\dagger}_{\bm{k}}\tau^6M_{\bm{k}}&=\tau^6, \\
M^{\dagger}_{\bm{k}}D_{\bm{k}}M_{\bm{k}}&=\tilde{\epsilon}_{\bm{k}},
\end{align}
where the Bogoliubov rotation matrix takes the block form
\begin{equation}
M_{\bm{k}}=\begin{pmatrix}
		U_{\bm{k}} & X_{\bm{k}} \\
		V_{\bm{k}} & Y_{\bm{k}}
		\end{pmatrix}.
\end{equation}
Furthermore,
\begin{equation}
\tau^6 = \begin{pmatrix}
			\mathds{1}_{6} & \bm{0} \\
			\bm{0} & -\mathds{1}_{6}
			\end{pmatrix},
\end{equation}
and 
\begin{equation}
\tilde{\epsilon}_{\bm{k}}=\begin{pmatrix}
	\mathcal{E}_{\bm{k}\uparrow} & \bm{0} \\
	\bm{0} & \mathcal{E}_{-\bm{k}\downarrow}
	\end{pmatrix}
\end{equation}

\noindent is a $12\times 12$ diagonal matrix representing the bosonic eigenenergies for up-spins at momentum $\bm{k}$ and down-spins at momentum $-\bm{k}$, where the $6\times 6$ diagonal matrix $\mathcal{E}_{\bm{p}\alpha}$ carries the bosonic eigenenergies $\epsilon^s_{\bm{p}\alpha}$ along its diagonal with $s$ the band index whose values range in $\{u,v,w,x,y,z\}$, the six bands comprising our unit cell as illustrated in Fig.~\ref{Fig:unitCell}, while $\bm{p}$ and $\alpha$ are the momentum and spin polarization, respectively, and $\bm{0}$ is the $6\times 6$ zero matrix. It is to be noted here that $\tilde{\epsilon}_{\bm{k}}$ has this form due to $SU(2)$ symmetry. Note, however, that for chiral ans\"atze one has $\epsilon^s_{\bm{k}\uparrow}\neq\epsilon^s_{-\bm{k}\downarrow}$, because the $\bm{k}\rightarrow-\bm{k}$ symmetry is broken. Nevertheless, we still have $\epsilon^s_{\bm{k}\uparrow}=\epsilon^s_{\bm{k}\downarrow}$ due to $SU(2)$ symmetry.

\section{Self-consistent mean-field parameters}\label{sec:self}

\begin{table}[]
\centering
 \caption{Self-consistent mean-field parameters and corresponding free energy per site $f_{\text{MF}}=\mathcal{F}_{\text{MF}}/N_s$ for the \emph{cuboc1},  \emph{cuboc2},  \emph{octahedral},  \emph{q=0}, and $\sqrt{3}\times\sqrt{3}$ SBMFT ans\"atze on the AFKM. The asterisk denotes a free mean-field parameter of the ansatz, which is self-consistently determined in the gapped spin liquid phase for spin $\mathcal{S}=0.2$.}
 \medskip
\begin{tabular}{|c"c|c|c|c|c}%{@{}|l|lllll@{}}
\bottomrule
\thickhline
 & \multicolumn{1}{c|}{ \emph{cuboc1}} & \multicolumn{1}{c|}{ \emph{cuboc2}} & \multicolumn{1}{c|}{ \emph{octahedral}} & \multicolumn{1}{c|}{ \emph{q=0}} & \multicolumn{1}{c|}{$\sqrt{3}\times\sqrt{3}$} \\ \midrule
\thickhline
$p1$ &  \multicolumn{1}{c|}{$1$} &  \multicolumn{1}{c|}{$1$} &  \multicolumn{1}{c|}{$1$} &  \multicolumn{1}{c|}{$0$} & \multicolumn{1}{c|}{$0$}  \\ \cmidrule(r){1-1}
\hline
 $\mathcal{A}$ & \multicolumn{1}{c|}{$0.2616$*} & \multicolumn{1}{c|}{$0.2624$*} & \multicolumn{1}{c|}{$0.2617$*} & \multicolumn{1}{c|}{$0.2626$*} & \multicolumn{1}{c|}{$0.2637$*}  \\ \cmidrule(r){1-1}
 \hline
 $\phi_{\mathcal{A}'}$ & \multicolumn{1}{c|}{$1.0143$*} & \multicolumn{1}{c|}{0} & \multicolumn{1}{c|}{$\pi$} & \multicolumn{1}{c|}{$0$} & \multicolumn{1}{c|}{$\pi$}  \\ \cmidrule(r){1-1}
 \hline
 $\mathcal{B}$ & \multicolumn{1}{c|}{$0.0540$*} & \multicolumn{1}{c|}{$0.0535$*} & \multicolumn{1}{c|}{$0.0536$*} & \multicolumn{1}{c|}{$0.0577$*} & \multicolumn{1}{c|}{$0.0574$*} \\ \cmidrule(r){1-1}
 \hline
$\phi_{\mathcal{B}}$  & \multicolumn{1}{c|}{$\pi$} & \multicolumn{1}{c|}{$3.1417$*} & \multicolumn{1}{c|}{$3.1416$*} & \multicolumn{1}{c|}{$\pi$} & \multicolumn{1}{c|}{$\pi$} \\ \cmidrule(r){1-1}
 \hline
$\phi_{\mathcal{B}'}$ & \multicolumn{1}{c|}{$\pi$} &\multicolumn{1}{c|}{ $-\phi_{\mathcal{B}}$} & \multicolumn{1}{c|}{$\phi_{\mathcal{B}}$} &\multicolumn{1}{c|}{$\pi$} & \multicolumn{1}{c|}{$\pi$} \\ \cmidrule(r){1-1}
 \hline
 $\lambda$ & \multicolumn{1}{c|}{$0.4086$*} & \multicolumn{1}{c|}{$0.4137$*} & \multicolumn{1}{c|}{$0.4096$*} & \multicolumn{1}{c|}{$0.4125$*} & \multicolumn{1}{c|}{$0.4182$*} \\ \cmidrule(r){1-1}
 \thickhline
  $f_{\text{MF}}$ & \multicolumn{1}{c|}{$-0.13127$} &\multicolumn{1}{c|}{ $-0.13200$} & \multicolumn{1}{c|}{$-0.13134$} &\multicolumn{1}{c|}{$-0.13148$} & \multicolumn{1}{c|}{$-0.13266$} \\ \cmidrule(l){6-6}
\hline

\hline
\thickhline
\end{tabular}
\label{Table}
\end{table}

\begin{figure*}
\begin{center}
\includegraphics[width= \textwidth]{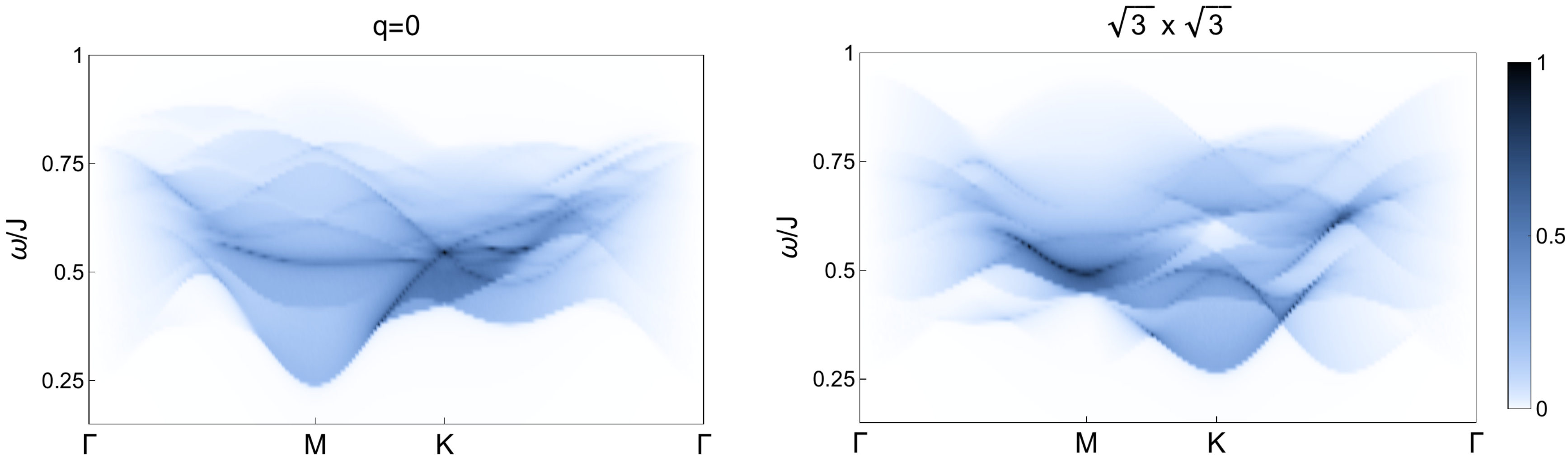}
\caption{The normalized dynamic structure factor along the $\Gamma$-M-K-$\Gamma$ high-symmetry lines for the non-chiral SBMFT ans\"atze \emph{q=0} (left) and $\sqrt{3}\times\sqrt{3}$ (right) in the gapped spin liquid phase at $\mathcal{S}=0.2$. }
\label{fig:nonchiral}
\end{center}
\end{figure*}

Before proceeding with the numerical computation of the spin structure factors we first determine the self-consistent mean-field parameters for each ansatz that we consider by extremizing the free energy
\begin{align}
\frac{\partial{\mathcal{F}_{\text{MF}}}}{\partial\mathcal{O}_j}&=0, \\
\frac{\partial{\mathcal{F}_{\text{MF}}}}{\partial\lambda}&=0,
\end{align}
where $\mathcal{O}_j$ are the free mean-field parameters (denoted by an asterisk in Table \ref{Table}) of the given ansatz, and $\mathcal{F}_{\text{MF}}$ is the mean-field free energy derived from Eq.~\eqref{eq:MFH} to be
\begin{align}
\nonumber \mathcal{F}_{\text{MF}}=\sum_{\bm{k}}^{\text{B.z.}}\sum_s\epsilon^s_{\bm{k}\uparrow}&+2JN_s(|\mathcal{A}|^2-|\mathcal{B}|^2) \\
&-\lambda N_s(1+2\mathcal{S}),
\end{align}
where B.z. stands for the 1st Brillouin zone. In this work we consider two non-chiral ans\"atze (\emph{q=0} and $\sqrt{3}\times\sqrt{3}$) as well as three further ans\"atze (\emph{cuboc1}, \emph{cuboc2}, and \emph{octahedral}) that can break time-reversal.\cite{Lhuillier2012,Misguich2013} In the following we set $J=1$ and $\mathcal{S}=0.2$. With this artificially small value of the spin we ensure that all ans\"atze describe a state deep in the spin-liquid phase, which is what we're interested in. We find the stationary point of $\mathcal{F}_{\text{MF}}$ by an adaptive-grid method that seeks to minimize $\sum_j(\partial{\mathcal{F}_{\text{MF}}}/\partial{\mathcal{O}}_j)^2$, stopping only when this sum is of the order of $10^{-8}$ or better. The results of this extremization procedure for all ans\"atze that we consider are shown in Table \ref{Table}. The self-consistent mean-field parameters for the different ans\"atze turn out to be quite close to one another. In particular, all five ans\"atze exhibit very similar values for $\mathcal{A}$, $\mathcal{B}$, and $\lambda$, the only parameters that are free in all ans\"atze. The main difference is in the phases, some of which are fixed by the specific form of an ansatz, while others are free.

It is important to note that the phase $\phi_\mathcal{B}$ for the \emph{cuboc2} as well as the \emph{octahedral} ansatz turns out to be equal to $\pi$ within numerical accuracy. Consequently, the saddle points of these two ans\"atze describe non-chiral spin-liquid phases, where time-reversal and parity symmetries are restored. In our computation only the \emph{cuboc1} ansatz turns out to be chiral. Note, however, that interactions beyond nearest neighbors can stabilize chiral saddle points of the \emph{cuboc2} form, where the phase $\phi_\mathcal{B}$ takes a non-trivial value.\cite{Lhuillier2012}

\section{Spin structure factors}\label{sec:ssf}
The dynamic spin structure factor is defined as
\begin{equation}\label{eq:spinstructurefactor}
S(\bm{k},\omega)=\frac{1}{N_s}\sum_{l,j}e^{\text{i}\bm{k}\cdot(\bm{r}_l-\bm{r}_j)}\int_{-\infty}^{\infty}dt e^{-\text{i}\omega t} \langle\mathbf{\hat{S}}_l(t)\cdot\mathbf{\hat{S}}_j\rangle,
\end{equation}
which, using the Bogoliubov operators, can be expressed at $T=0$ and in the absence of a spinon condensate as
\begin{widetext}
\begin{align}\label{eq:dsf}
\nonumber S(\bm{k},\omega)%=&\frac{9}{N_s}\int_{-\infty}^{\infty}dt e^{-\text{i}\omega t}\sum_{s,r,m,n}\sum_{\bm{q}}^{\text{B.z.}}\bigg\{\\ \nonumber
%&\hat{X}^*_{sn}(\bm{q})\hat{U}_{sm}(\bm{k}+\bm{q})\bigg[\hat{U}^*_{rm}(\bm{k}+\bm{q})\hat{X}_{rn}(\bm{q})-\hat{Y}_{rn}(\bm{q})\hat{V}^*_{rm}(\bm{k}+\bm{q})\bigg]\langle\hat{\gamma}^n_{-\bm{q}\downarrow}(t)\hat{\gamma}^{n\dagger}_{-\bm{q}\downarrow}\rangle\langle\hat{\gamma}^m_{\bm{k}+\bm{q}\uparrow}(t)\hat{\gamma}^{m\dagger}_{\bm{k}+\bm{q}\uparrow}\rangle+\\ \nonumber
%&2\hat{X}^*_{sn}(\bm{q})\hat{Y}^*_{sm}(-\bm{k}-\bm{q})\bigg[\hat{Y}_{rm}(-\bm{k}-\bm{q})\hat{X}_{rn}(\bm{q})+\hat{Y}_{rn}(\bm{q})\hat{X}_{rm}(-\bm{k}-\bm{q})\bigg]\langle\hat{\gamma}^n_{-\bm{q}\downarrow}(t)\hat{\gamma}^{n\dagger}_{-\bm{q}\downarrow}\rangle\langle\hat{\gamma}^m_{\bm{k}+\bm{q}\downarrow}(t)\hat{\gamma}^{m\dagger}_{\bm{k}+\bm{q}\downarrow}\rangle+\\ \nonumber 
%&2\hat{V}_{sn}(-\bm{q})\hat{U}_{sm}(\bm{k}+\bm{q})\bigg[\hat{U}^*_{rm}(\bm{k}+\bm{q})\hat{V}^*_{rn}(-\bm{q})+\hat{U}^*_{rn}(-\bm{q})\hat{V}^*_{rm}(\bm{k}+\bm{q})\bigg]\langle\hat{\gamma}^n_{-\bm{q}\uparrow}(t)\hat{\gamma}^{n\dagger}_{-\bm{q}\uparrow}\rangle\langle\hat{\gamma}^m_{\bm{k}+\bm{q}\uparrow}(t)\hat{\gamma}^{m\dagger}_{\bm{k}+\bm{q}\uparrow}\rangle+\\ \nonumber  
%&\hat{V}_{sn}(-\bm{q})\hat{Y}^*_{sm}(-\bm{k}-\bm{q})\bigg[\hat{Y}_{rm}(-\bm{k}-\bm{q})\hat{V}^*_{rn}(-\bm{q})-\hat{U}^*_{rn}(-\bm{q})\hat{X}_{rm}(-\bm{k}-\bm{q})\bigg]\langle\hat{\gamma}^n_{-\bm{q}\uparrow}(t)\hat{\gamma}^{n\dagger}_{-\bm{q}\uparrow}\rangle\langle\hat{\gamma}^m_{\bm{k}+\bm{q}\downarrow}(t)\hat{\gamma}^{m\dagger}_{\bm{k}+\bm{q}\downarrow}\rangle \bigg\} \\ \nonumber
=&\frac{3}{2}\sum_{s,r,m,n}\frac{1}{N_q}\sum_{\bm{q}}^{\text{B.z.}}\bigg\{\\ \nonumber
&X^*_{sn}(\bm{-q})U_{sm}(\bm{k}-\bm{q})\bigg[U^*_{rm}(\bm{k}-\bm{q})X_{rn}(-\bm{q})-Y_{rn}(-\bm{q})V^*_{rm}(\bm{k}-\bm{q})\bigg]\delta(\omega-\epsilon^n_{\bm{q}\downarrow}-\epsilon^m_{\bm{k}-\bm{q}\uparrow})+\\ \nonumber
&2X^*_{sn}(-\bm{q})Y^*_{sm}(-\bm{k}+\bm{q})\bigg[Y_{rm}(-\bm{k}+\bm{q})X_{rn}(-\bm{q})+Y_{rn}(-\bm{q})X_{rm}(-\bm{k}+\bm{q})\bigg]\delta(\omega-\epsilon^n_{\bm{q}\downarrow}-\epsilon^m_{\bm{k}-\bm{q}\downarrow})+\\ \nonumber 
&2V_{sn}(\bm{q})U_{sm}(\bm{k}-\bm{q})\bigg[U^*_{rm}(\bm{k}-\bm{q})V^*_{rn}(\bm{q})+U^*_{rn}(\bm{q})V^*_{rm}(\bm{k}-\bm{q})\bigg]\delta(\omega-\epsilon^n_{\bm{q}\uparrow}-\epsilon^m_{\bm{k}-\bm{q}\uparrow})+\\ 
&V_{sn}(\bm{q})Y^*_{sm}(-\bm{k}+\bm{q})\bigg[Y_{rm}(-\bm{k}+\bm{q})V^*_{rn}(\bm{q})-U^*_{rn}(\bm{q})X_{rm}(-\bm{k}+\bm{q})\bigg]\delta(\omega-\epsilon^n_{\bm{q}\uparrow}-\epsilon^m_{\bm{k}-\bm{q}\downarrow}) \bigg\},
\end{align}
\end{widetext}
where $s$, $r$, $m$, and $n$ are band indices taking values in $\{u,v,w,x,y,z\}$. The static (equal-time) structure factor is obtained by integrating over frequencies $S(\bm{k}) = \int d\omega S(\bm{k},\omega)$.

\begin{figure*}
\begin{center}
\includegraphics[width= \textwidth]{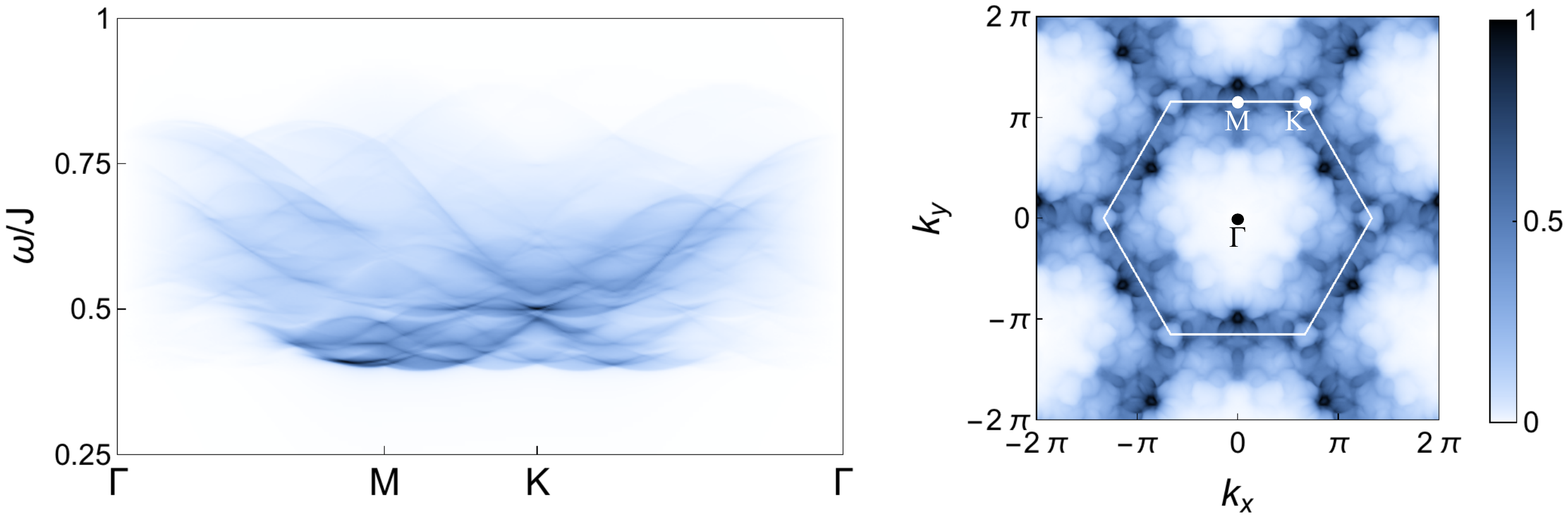}
\caption{The normalized dynamic structure factor for the chiral \emph{cuboc1} ansatz along the $\Gamma$-M-K-$\Gamma$ high-symmetry lines (left) and in the $\bm{k}$-plane at $\omega=0.45J$ (right) in the gapped spin liquid phase at $\mathcal{S}=0.2$. The white hexagon in the right panel marks the extended Brillouin zone. Note that the dynamic structure factor at fixed frequency (right) is not symmetric under inversion and only has a three-fold rotation symmetry due to time-reversal symmetry breaking (see main text).}
\label{fig:cuboc1}
\end{center}
\end{figure*}

\section{Results and discussion}\label{sec:results}

We use the VEGAS\cite{Lepage1978} Monte Carlo integration routine to numerically evaluate the dynamic structure factors of the ans\"atze shown in Table~\ref{Table}, while approximating the Dirac $\delta$-functions in Eq.~\eqref{eq:dsf} as Lorentzian functions with a width $10^{-3}$ for numerical reasons.

\begin{figure*}
\begin{center}
\includegraphics[width= \textwidth]{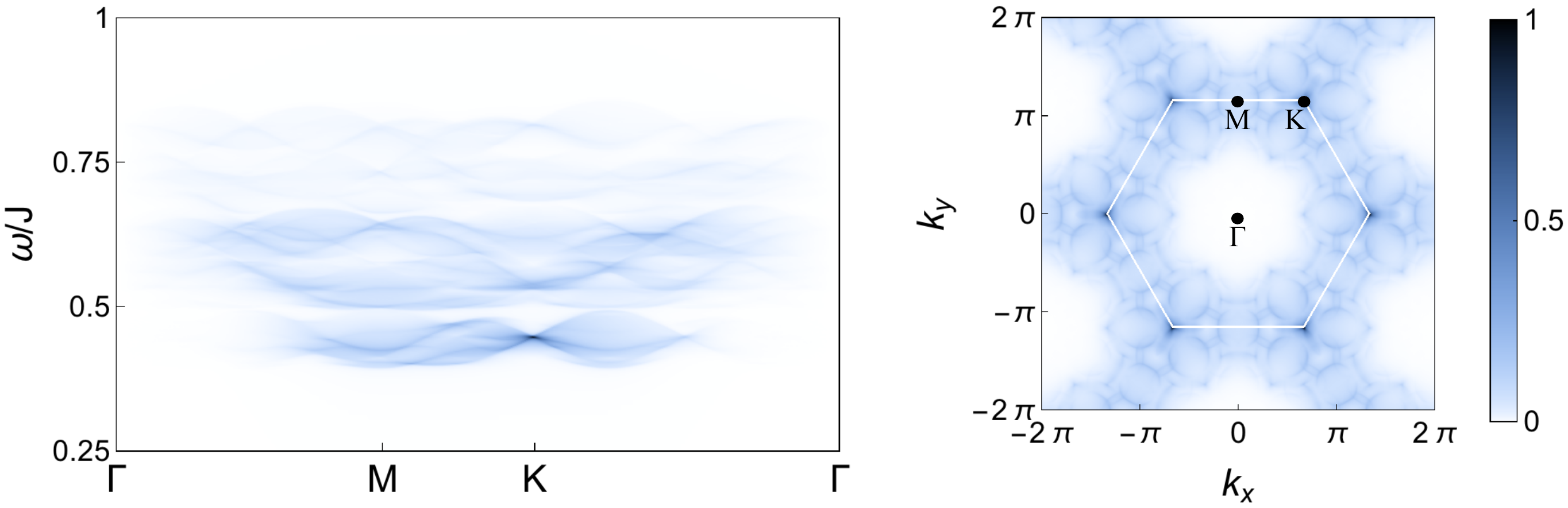}
\caption{The normalized dynamic structure factor for the \emph{cuboc2} ansatz along the $\Gamma$-M-K-$\Gamma$ high-symmetry lines (left) and in the $\bm{k}$-plane at $\omega=0.45J$ (right) in the gapped spin liquid phase at $\mathcal{S}=0.2$. 
Note that the saddle point values of the mean-field parameters for the \emph{cuboc2} ansatz preserve time-reversal symmetry at $\mathcal{S}=0.2$, consequently this state is not chiral.}
\label{fig:cuboc2}
\end{center}
\end{figure*}

\begin{figure*}
\begin{center}
\includegraphics[width= \textwidth]{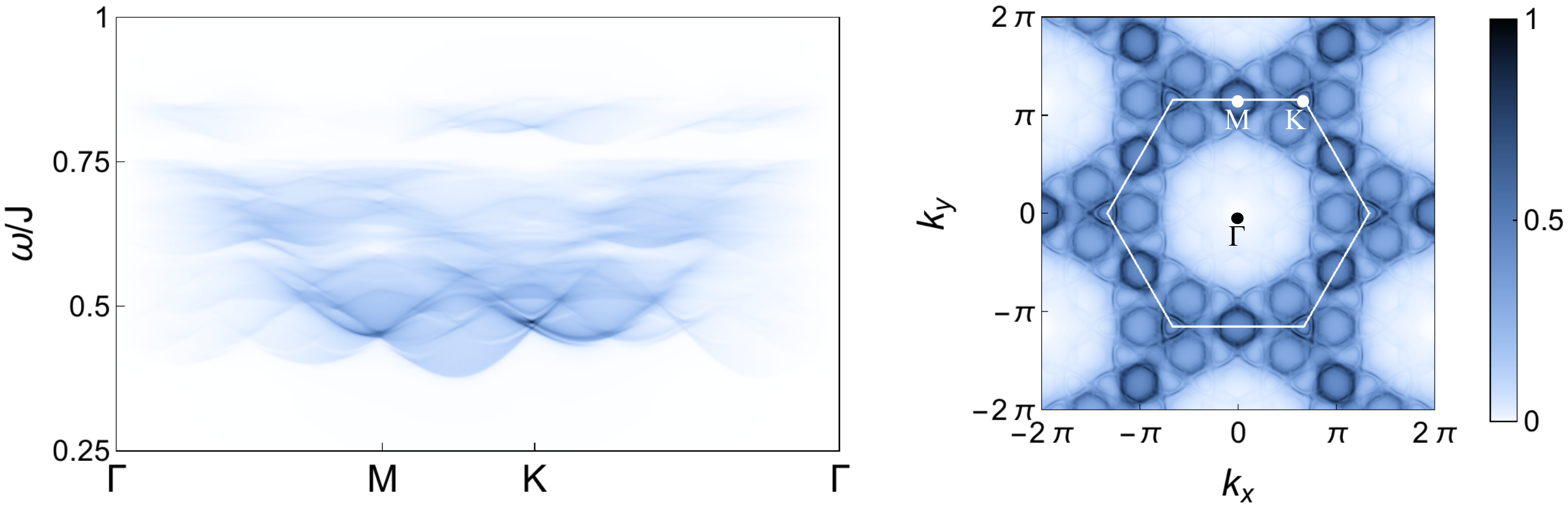}
\caption{The normalized dynamic structure factor for the  \emph{octahedral} ansatz along the $\Gamma$-M-K-$\Gamma$ high-symmetry lines (left) and in the $\bm{k}$-plane at $\omega=0.48J$ (right) in the gapped spin liquid phase at $\mathcal{S}=0.2$. Note that the saddle point is non-chiral, as in the case of the \emph{cuboc2} ansatz.}
\label{fig:octahedral}
\end{center}
\end{figure*}

The non-chiral ans\"atze  \emph{q=0} and $\sqrt{3}\times\sqrt{3}$ were first discussed in Ref.~[\onlinecite{Sachdev1992}], and their dynamic structure factors were calculated in Ref.~[\onlinecite{Punk2014}], although using an ansatz with $\langle\mathcal{B}_{ij}\rangle=0$. Similar dynamical structure factors for fermionic mean-field spin liquids have been computed in Ref.~[\onlinecite{Dodds}]. In Fig.~\ref{fig:nonchiral}, we show their dynamic structure factors along the $\Gamma$-M-K-$\Gamma$ high-symmetry lines in the gapped spin liquid phase with $\mathcal{S}=0.2$, using the self-consistent mean-field parameters shown in Table~\ref{Table}. Note that we've adopted a normalization where the maximum of the structure factors is set to unity for convenience. Our results are qualitatively very similar to those in Ref.~[\onlinecite{Punk2014}]. The small differences come from the fact that the spinon dispersions are slightly altered when taking non-zero $\langle \mathcal{B}_{ij} \rangle$'s into account. 

\begin{figure*}
\begin{center}
\includegraphics[width= \textwidth]{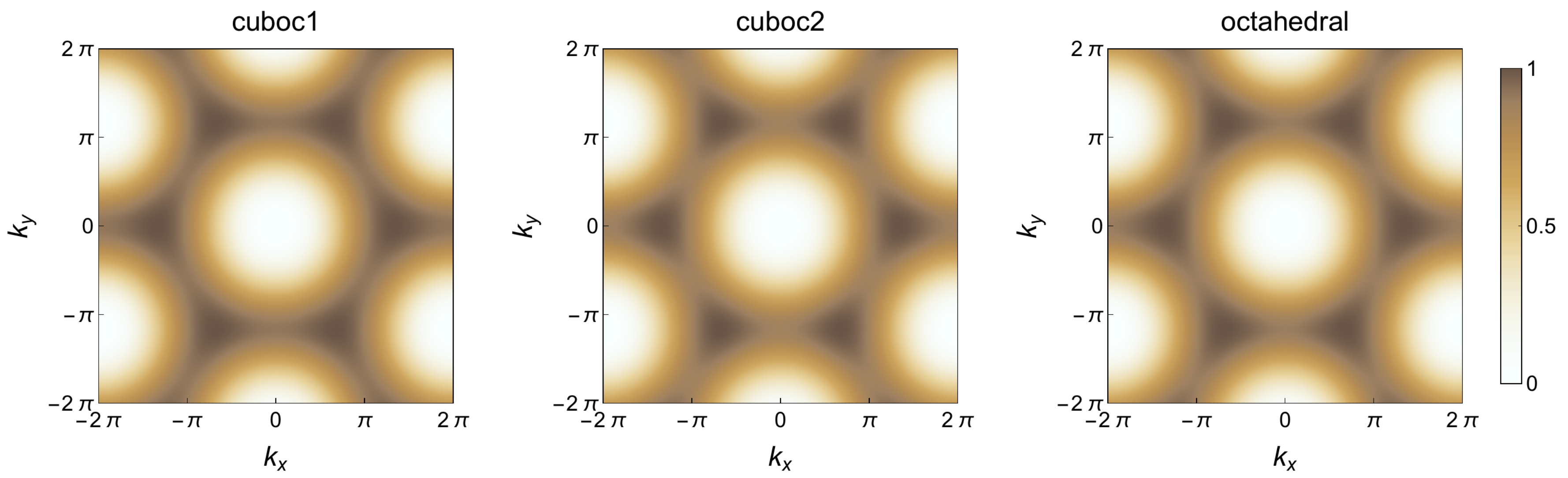}
\caption{(Color online) Normalized static spin structure factors for the \emph{cuboc1}, \emph{cuboc2}, and \emph{octahedral} ans\"atze in the gapped spin liquid phase with spin $\mathcal{S}=0.2$. Note that the static structure factor of the chiral  \emph{cuboc1} state does not show signs of time-reversal symmetry breaking (see main text).}
\label{fig:ssf}
\end{center}
\end{figure*}

Figs.~\ref{fig:cuboc1},~\ref{fig:cuboc2}, and~\ref{fig:octahedral} show the dynamic structure factors in the $\bm{k}$-plane for fixed frequencies $\omega$, as well as along the $\Gamma$-M-K-$\Gamma$ high-symmetry lines for the  \emph{cuboc1}, \emph{cuboc2}, and \emph{octahedral} ans\"atze, respectively, in the gapped spin liquid phase with spin $\mathcal{S}=0.2$. The dynamic structure factor at $\omega=0.45J$ for \emph{cuboc1} in Fig.~\ref{fig:cuboc1} shows that inversion symmetry is lost with respect to the $\Gamma$-point due to time-reversal symmetry breaking. Consequently, the usual six-fold rotational symmetry is reduced to a three-fold one. One would expect to see the same for the \emph{cuboc2} (at $\omega=0.45J$) and  \emph{octahedral} (at $\omega=0.48J$) ans\"atze in Figs.~\ref{fig:cuboc2} and~\ref{fig:octahedral}, respectively, as they allow for nontrivial Aharonov-Bohm phases when a spinon is taken around a plaquette. However, as shown in Table~\ref{Table}, the saddle point value of $\phi_\mathcal{B}$ at $\mathcal{S}=0.2$ is equal to $\pi$ within numerical accuracy, which, along with the fixed phase $\phi_{\mathcal{A}'}=0$ (\emph{cuboc2}) or $\pi$ (\emph{octahedral}), leads to a time-reversal invariant ansatz. The $\bm{k}\rightarrow-\bm{k}$ symmetry as well as the six-fold rotational symmetry is thus retained in the dynamical structure factor. On the other hand, for the  \emph{cuboc1} ansatz, the free phase $\phi_{\mathcal{A}'}$ takes on a value other than $n\pi$ ($n\in\mathbb{Z}$), leading to a chiral spin liquid with broken time-reversal and parity symmetry. An interesting feature in the \emph{cuboc1} dynamical structure factor along the $\Gamma$-M-K-$\Gamma$ high-symmetry lines in Fig.~\ref{fig:cuboc1} is the relatively flat onset of the two-spinon continuum compared to the \emph{q=0} and $\sqrt{3} \times \sqrt{3}$ case.

 The static structure factors for the  \emph{cuboc1},  \emph{cuboc2}, and  \emph{octahedral} ans\"atze, shown in Fig.~\ref{fig:ssf}, are qualitatively very similar. 
Note that the static structure factor for the chiral \emph{cuboc1} state doesn't show signs of time-reversal or parity symmetry breaking. This can be understood by recognizing from Eq.~\eqref{eq:spinstructurefactor} that one can write
\begin{align}\label{eq:spinstructurefactor-k}\nonumber
S(-\bm{k},\omega)&=\frac{1}{N_s}\sum_{l,j}e^{\text{i}\bm{k}\cdot(\bm{r}_l-\bm{r}_j)}\int_{-\infty}^{\infty}dt e^{-\text{i}\omega t} \langle\mathbf{\hat{S}}_l\cdot\mathbf{\hat{S}}_j(t)\rangle\\
&=\frac{1}{N_s}\sum_{l,j}e^{\text{i}\bm{k}\cdot(\bm{r}_l-\bm{r}_j)}\int_{-\infty}^{\infty}dt e^{-\text{i}\omega t} \langle\mathbf{\hat{S}}_l(-t)\cdot\mathbf{\hat{S}}_j\rangle
\end{align}
by a simple relabeling of the site indices. It is clear from Eqs.~\eqref{eq:spinstructurefactor-k} and~\eqref{eq:spinstructurefactor} that $S(-\bm{k},\omega)=S(\bm{k},\omega)$ only if $\langle\mathbf{\hat{S}}_l(-t)\cdot\mathbf{\hat{S}}_j\rangle=\langle\mathbf{\hat{S}}_l(t)\cdot\mathbf{\hat{S}}_j\rangle$ is time reversal invariant. On the other hand, the static (equal-time) structure factor is always invariant under $\bm{k} \to -\bm{k}$.

Lastly, the variational groundstate energies of the five different ans\"atze considered here are listed in the last line of Table \ref{Table}. We find that the non-chiral $\sqrt{3} \times \sqrt{3}$ ansatz has the lowest energy at $\mathcal{S}=0.2$. Note that this is in contrast to Ref. [\onlinecite{Lhuillier2012}], who find that the \emph{cuboc1} state has the lowest energy. It is worth mentioning here that the SBMFT approach is not quantitatively reliable to find the true groundstate of the kagome Heisenberg antiferromagnet, nor does it give variational upper bounds to the groundstate energy. This is due to the fact that the spin-length constraint is only imposed on average, and thus unphysical states are only approximately projected out. Consequently, the main purpose of our work is not to determine the true groundstate of the kagome Heisenberg antiferromagnet, but to highlight features in dynamical correlation functions of different chiral spin liquid states.

\section{Conclusion}\label{sec:conclusion}

We computed static and dynamic spin structure factors of several chiral and non-chiral SBMFT ans\"atze deep in the gapped spin liquid phase at spin $\mathcal{S}=0.2$. Even though the \emph{cuboc1}, \emph{cuboc2}, and \emph{octahedral} ans\"atze allow for time-reversal symmetry breaking, only the saddle-point of the \emph{cuboc1} ansatz is chiral, which can be seen directly in the dynamic structure factor at fixed frequency, as seen in Fig.~\ref{fig:cuboc1}. Time-reversal symmetry breaking manifests itself by breaking the inversion symmetry with respect to the $\Gamma$-point, as well as reducing the usual six-fold rotational symmetry present for non-chiral ans\"atze to a three-fold rotational symmetry. 

\acknowledgements

J.C.H. is grateful to Pavel Kos and Lode Pollet for fruitful discussions, as well as to Guy Almog for providing Fig.~\ref{Fig:unitCell}. This work is supported by the Nano-Initiative Munich (NIM).

%----------------------------------------------
%\begin{thebibliography}{99}
%*********************************************************************************************

%--------------------------------------------------------
\end{document}